\documentclass[final,5p,11pt]{elsarticle}
\pdfoutput=1
\usepackage[utf8x]{inputenc}  
\usepackage[T1]{fontenc}      
\usepackage{amsmath,amssymb}
\usepackage{booktabs,tabularx}
\usepackage[absolute]{textpos}
\usepackage{xspace}
\usepackage{times}
\usepackage[bookmarks=true,linktocpage]{hyperref}
\hypersetup{pdftitle={E6 model benchmarks},
            pdfauthor={P. Athron, D. Harries, R. Nevzorov, A.G. Williams},
            pdfkeywords={supersymmetry,E6,dark matter}}

\biboptions{numbers,sort&compress}

\newcommand{\FS}{FlexibleSUSY\@\xspace}
\newcommand{\SOFTSUSY}{SOFTSUSY\@\xspace}
\newcommand{\FSv}{FlexibleSUSY-1.1.0\@\xspace} 

\newcommand{\SARAHv}{SARAH-4.5.6\@\xspace} 

\newcommand{\SUSYHDv}{SUSYHD-1.0.2\@\xspace}

\newcommand{\MICROMEGASv}{micrOMEGAs-4.1.8\@\xspace}
\newcommand{\CalcHEP}{CalcHEP\@\xspace}
\newcommand{\xenon}{XENON1T\@\xspace}
\newcommand{\textoverline}[1]{$\overline{\mbox{#1}}$}
\newcommand{\DRbar}{\textoverline{DR}\xspace}
\newcommand{\secref}[1]{Section~\ref{#1}}
\newcommand{\tabref}[1]{\tablename~\ref{#1}}

\journal{Physics Letters B}

\begin{document}

\begin{frontmatter}

\title{$E_6$ Inspired SUSY Benchmarks, Dark Matter Relic Density and a $125$ GeV
Higgs}

\author[monash]{Peter Athron}
\author[adelaide]{Dylan Harries\corref{cor1}}
\author[adelaide]{Roman Nevzorov}
\author[adelaide]{Anthony G. Williams}

\cortext[cor1]{Corresponding author,\\\texttt{dylan.harries@adelaide.edu.au}}

\address[monash]{ARC Centre of Excellence for Particle Physics at
the Terascale, School of Physics, Monash University, Melbourne, Victoria 3800,
Australia}
\address[adelaide]{ARC Centre of Excellence for Particle Physics at
the Terascale, Department of Physics, The University of Adelaide, Adelaide,
South Australia 5005, Australia}

\begin{abstract}
We explore the relic density of dark matter and the particle spectrum
within a constrained version of an $E_6$ inspired SUSY model with an
extra $U(1)_N$ gauge symmetry.  In this model a single exact custodial
symmetry forbids tree-level flavor-changing transitions and the most
dangerous baryon and lepton number violating operators.  We present a
set of benchmark points showing scenarios that have a SM-like Higgs
mass of 125 GeV and sparticle masses above the LHC limits.  They lead
to striking new physics signatures which may be observed during run II
of the LHC and can distinguish this model from the simplest SUSY
extensions of the SM.  At the same time these benchmark scenarios are
consistent with the measured dark matter abundance and necessarily
lead to large dark matter direct detection cross sections close to current
limits and observable soon at the \xenon experiment.
\end{abstract}


\end{frontmatter}

\begin{textblock*}{7em}(0.95\textwidth,1cm)
\noindent\footnotesize
ADP--15--53/T955 \\
CoEPP--MN--15--14 \\
\end{textblock*}

\section{Introduction}
\noindent With the discovery of the 125 GeV Higgs boson
\cite{Aad:2012tfa,Chatrchyan:2012xdj} made in run I of the Large
Hadron Collider (LHC), the primary goal of run II of the LHC is now to
look for signs of physics beyond the standard model (SM).  The best
motivated class of extensions of the SM are models based on
low--energy supersymmetry (SUSY).  Supersymmetry is the most general
extension of the Poincar\'{e} group \cite{Coleman:1967ad,Haag:1974qh}.
When the new SUSY partners have masses around the TeV scale the
minimal supersymmetric standard model (MSSM) allows to address the
hierarchy problem, to achieve the unification of the SM gauge
couplings, allowing the MSSM to be embedded into a Grand Unified
Theory (GUT), and to predict the correct relic abundance of dark
matter (DM) simultaneously.

$E_6$ inspired SUSY models provide a very
attractive framework for GUT scale physics and can arise from $E_8
\times E_8^\prime$ heterotic string theory
\cite{delAguila:1985cb,Kaplunovsky:1993rd,Brignole:1993dj}.  At low
energies these models can lead to an extra $U(1)$ gauge symmetry which
is spontaneously broken, giving rise to an effective $\mu$ term and
a massive $Z^\prime$ gauge boson.
$E_6$ inspired SUSY extensions of the SM gathered a lot of attention in
the past (see, for example, \cite{Hewett:1988xc,Gunion:1989we,Binetruy:1985xm,
  Ellis:1986yg,Ibanez:1986si,Gunion:1986ky,Haber:1986gz,Ellis:1986ip,
  Drees:1987tp,Baer:1987eb,Gunion:1987jd}).

More recently the exceptional supersymmetric standard model (E$_6$SSM)
was proposed \cite{King:2005jy,King:2005my,Athron:2008np,Athron:2010zz} where
right--handed neutrinos have zero charge under the extra $U(1)_{N}$
gauge symmetry.  Only in this case can the right--handed neutrinos be
superheavy, allowing the see-saw mechanism to explain the mass hierarchy
in the lepton sector and providing a mechanism for the generation of
the baryon asymmetry in the Universe via leptogenesis
\cite{King:2008qb}.  Different modifications of this SUSY model were
also considered \cite{Howl:2007hq,Howl:2007zi,Nevzorov:2012hs}.

To obtain realistic phenomenology the E$_6$SSM has an
approximate $Z_2^H$ symmetry to forbid large flavor-changing neutral
currents (FCNCs), as well as another exact $Z_2$ symmetry which plays a
similar role to $R$-parity in the MSSM.  The existence of light exotic states
in this model, which are not present in the MSSM, could explain the observed
relic DM density \cite{Hall:2009aj}.  However such scenarios
also imply that the lightest SM--like Higgs boson decays predominantly into
DM exotic states, which also have an unacceptably large spin independent elastic
cross section  \cite{Hall:2010ix}.  Thus the corresponding scenarios have been
ruled out by DM direct detection and LHC experiments.  The proposed
phenomenologically viable modification of the E$_6$SSM requires the imposition
of another discrete symmetry \cite{Hall:2011zq} in addition to the set of
approximate and exact discrete symmetries mentioned above, to prevent an
MSSM-like neutralino from decaying into these exotic states.

Here we investigate for the first time the constrained version of a
recently proposed alternative modification of the E$_6$SSM (CSE$_6$SSM)
\cite{Athron:2014pua}.  This model makes use of recent work on $E_6$ orbifold
GUTs \cite{Nevzorov:2012hs} where an exact discrete symmetry was found which
forbids both couplings that induce large FCNCs and those that lead to rapid
proton decay.  At the same time the model also conserves matter parity, which
implies that there is a bino-like or Higgsino-like stable DM candidate.  In
fact the CSE$_6$SSM has two potential DM candidates, as the discrete symmetry
which forbids FCNCs and proton decay leads to the lightest exotic particle also
being stable.

In this letter we demonstrate that a DM candidate stabilized
by the automatic conservation of matter parity is sufficient to fit
the relic DM density within the CSE$_6$SSM while the second
candidate is almost massless and therefore contributes negligibly to
the DM density in the simplest phenomenologically viable
scenarios.  In this way we can explain the measured density of DM, while
also satisfying LHC constraints such as the 125 GeV Higgs mass
measurement and mass limits on sparticles and exotic states.  We find
that some sparticles and new exotic states can be within reach of
run II of the LHC and that DM states have sufficiently large direct
detection cross-sections close to current limits and observable soon
at the \xenon experiment.  We present benchmark points showing scenarios
that could be discovered in the very near future in either of these
experiments and urgently need to be investigated.  This letter is
intended to be followed by a more detailed companion paper which will
give analytic expressions used; describe the methodology in detail;
provide a thorough exploration of the parameter space, with detailed
plots of the interesting regions and make a comparison to the MSSM.

Previously the electroweak symmetry breaking (EWSB) of $E_6$ models
with an extra $U(1)$ has been investigated
\cite{Suematsu:1994qm,Cvetic:1995rj,Cvetic:1996mf,
  Cvetic:1997ky,Keith:1997zb,Langacker:1998tc,Daikoku:2000ep} and a
mechanism for radiative EWSB demonstrated
\cite{Athron:2009bs,Athron:2009ue}.  These models can increase the
theoretical upper bound on the lightest Higgs boson mass
\cite{Daikoku:2000ep,
  Barger:2007ay,King:2005jy,King:2005my,King:2006vu,Accomando:2006ga,
  King:2006rh,Barger:2006dh}.  The renormalization of the vacuum
expectation values (VEVs) was considered in
Refs.~\cite{Sperling:2013eva,Sperling:2013xqa} and the impact of gauge
kinetic mixing when two extra $U(1)$ gauge groups are at low energies
was investigated \cite{Rizzo:2012rf}.  These models may ameliorate the
little hierarchy problem but have new contributions to fine tuning
from the $Z^\prime$ mass \cite{Athron:2013ipa,Athron:2015tsa}.  The
consequences for neutrino physics have been examined
\cite{Kang:2004ix,Ma:1995xk}, as well as leptogenesis
\cite{Hambye:2000bn,King:2008qb} and electroweak (EW) baryogenesis
\cite{Ma:2000jf,Kang:2004pp}.  There have been many studies into the
extended set of neutralinos \cite{Keith:1997zb,Suematsu:1997tv,
  GutierrezRodriguez:2006hb,Suematsu:1997qt,Suematsu:1997au,Keith:1996fv,
  Hesselbach:2001ri,Barger:2005hb,Choi:2006fz,Barger:2007nv,Gherghetta:1996yr,
  Barger:2007ay}.  The muon anomalous magnetic moment
\cite{Grifols:1986vr, Morris:1987fm}, electric dipole moments
\cite{Suematsu:1997tv, GutierrezRodriguez:2006hb}, $\mu\to e\gamma$
\cite{Suematsu:1997qt} and CP-violation in the Higgs sector
\cite{Ham:2008fx} have been investigated.  Anomaly mediated SUSY
breaking \cite{Asano:2008ju} and family symmetries
\cite{Stech:2008wd,Howl:2008xz, Howl:2009ds} have been studied in these
$U(1)$ extensions of the SM.

The signatures associated with the exotic states in these models
have been considered \cite{Kang:2007ib,Athron:2011wu} and $Z'$ mass limits
at the LHC and Tevatron were examined \cite{Accomando:2010fz}.  The
impact of the 125 GeV Higgs observation and LHC limits on sparticles
was examined \cite{Athron:2012sq} and was re-examined after calculating
higher order corrections to gauge and Yukawa couplings \cite{Athron:2012pw}.
Non-standard Higgs decays have also been studied \cite{Hall:2010ix,
  Nevzorov:2014sha,Athron:2014pua}.  What a measurement of the first and
second generation sfermion masses might tell us about the underlying $E_6$
GUT model was looked at \cite{Miller:2012vn}.  Finally the impact of gauge
kinetic mixing on $Z^\prime$ and slepton production at the LHC was examined
\cite{Krauss:2012ku}.

The structure of this letter is as follows.  In \secref{sec:model} the
model we investigate is described.  In \secref{sec:results} the
procedure used to investigate the model is explained and we describe
the results of our investigation.  We present benchmark scenarios
which fit current data, including the Higgs mass measurement and the
relic density of DM.  Finally in \secref{sec:conclusions} we give our
conclusions.

\section{The CSE$_6$SSM}
\label{sec:model}
\noindent Models with an extra $U(1)$ can arise from the breakdown of
$E_6$ GUTs.  Such GUT models can emerge from ten dimensional
$E_8 \times E_8^\prime$ heterotic string theory after the compactification
of extra dimensions, breaking $E_8 \rightarrow E_6$ \cite{delAguila:1985cb,
  Kaplunovsky:1993rd,Brignole:1993dj}.  The $E_8^\prime$ then forms a hidden
sector which interacts with the visible sector only through gravitational
interactions.  When local supergravity is broken in the hidden sector these
gravitational interactions transmit the SUSY breaking to the visible
sector, giving rise to a set of soft breaking masses.

If the $E_6$ gauge group lives in 5 or 6 dimensions then it may be broken by
the boundary conditions.  Five and six dimensional orbifold GUTs can then lead
to the $E_6$ inspired model with an exact custodial symmetry
\cite{Nevzorov:2012hs} and give rise to precisely the low energy model we
study in this letter, which we now describe in detail.

The low energy gauge group is that of the SM with an additional $U(1)_N$
symmetry.  This $U(1)_N$ is a linear combination of $U(1)_\psi$ and
$U(1)_{\chi}$,
\begin{equation}
U(1)_N = \frac{1}{4} U(1)_{\chi} + \frac{\sqrt{15}}{4} U(1)_{\psi} ,
\end{equation}
which appear in the breakdown of $E_6$ via $E_6 \to SO(10) \times U(1)_{\psi}$
and $SO(10) \to SU(5) \times U(1)_{\chi}$.

The matter content fills three complete generations of $E_6$
$\mathbf{27}$-plets, $27_i$, ensuring gauge anomalies automatically
cancel.  Each $27_i$ contains one generation of ordinary matter, a SM
singlet field $S_i$, up- and down-type Higgs doublets $H^{u}_{i}$ and
$H^{d}_{i}$\footnote{One pair of these doublets, $H_u$ and $H_d$,
  play the role of Higgs fields.  The other two generations of
  $H^{u}_{i}$ and $H^{d}_{i}$ are denoted ``inert Higgs'' since their
  scalar components don't develop VEVs.}  and charged $\pm 1/3$
leptoquarks $D_i$ and $\bar{D}_i$.  There are also two additional pairs
of states $(L_4,\overline{L}_4)$ and $(S,\overline{S})$ that originate
from $\mathbf{27}^\prime$ and $\overline{\mathbf{27}}^\prime$ and
automatically cancel anomalies on their own as a consequence.  This
structure of the low energy matter content allowing this cancellation
is not a coincidence, it is a consequence of the $E_6$ GUT, which is
anomaly free, and the specific orbifold GUT construction
\cite{Nevzorov:2012hs}.  The
representations and charges of the superfields are given in
\tabref{tab:charges}, where there and throughout this letter Roman
indices run over $i,j,k=1,2,3$ and Greek indices run over
$\alpha=1,2$.

As a consequence of the $E_6$ based construction, and the breaking of
the $U(1)_\chi$ and $U(1)_\psi$ at some intermediate scale, the model
automatically conserves matter parity, $Z_2^M =(-1)^{3(B-L)}$.
However there remain dangerous baryon number ($B$) and lepton number
($L$) violating interactions.  So to avoid rapid proton decay and FCNCs one
additional discrete symmetry $\tilde{Z}_2^H$ is imposed.  As a consequence
the model has not one, but two new stable particles.  This can be understood by
defining a $Z_2^E$ symmetry by $\tilde{Z}_2^H = Z_2^M \times Z_2^E$.
The charges under these discrete symmetries are specified in
\tabref{tab:charges}.  Since $\tilde{Z}_2^H$ and $Z_2^M$ are
separately conserved, $Z_2^E$ is also conserved.  In the cases studied
here this means that both the lightest exotic singlino associated with
the $\hat{S}_i$ superfields and the lightest ordinary neutralino are
stable.

\begin{table*}[ht]
\centering
\begin{tabular}{|c||c|c|c|c|c|c|c|c|c|c|c|c|c|c|c|c|c|}
 \hline
 & $\hat{Q}_i$ & $\hat{u}_i^c$ & $\hat{d}_i^c$ & $\hat{L}_i$ & $\hat{e}_i^c$
  & $\hat{N}_i^c$ & $\hat{S}$  & $\hat{\overline{S}}$ & $\hat{S}_i$
  & $\hat{H}_u$ & $\hat{H}_d$ & $\hat{H}_{\alpha}^u$ & $\hat{H}_{\alpha}^d$
  & $\hat{D}_i$ & $\hat{\overline{D}}$ & $\hat{L}_4$
  & $\hat{\overline{L}}_4$ \\
 \hline
 $SU(3)$ & $\bf{3}$ & $\bf{\overline{3}}$ &  $\bf{\overline{3}}$ & $\bf{1}$
  & $\bf{1}$ & $\bf{1}$ & $\bf{1}$ & $\bf{1}$ & $\bf{1}$ & $\bf{1}$ & $\bf{1}$
  & $\bf{1}$ & $\bf{1}$ &  $\bf{3}$ & $\bf{\overline{3}}$ & $\bf{1}$
  & $\bf{1}$ \\
 \hline
 $SU(2)$ & $\bf{2}$ & $\bf{1}$ &  $\bf{1}$ & $\bf{2}$ & $\bf{1}$ & $\bf{1}$
  & $\bf{1}$ & $\bf{1}$ & $\bf{1}$ & $\bf{2}$ & $\bf{2}$ & $\bf{2}$ & $\bf{2}$
  &  $\bf{1}$ & $\bf{1}$ & $\bf{2}$ & $\bf{\overline{2}}$ \\
 \hline
 $\sqrt{\frac{5}{3}}Q^{Y}_i$ & $\frac{1}{6}$ & $-\frac{2}{3}$ & $\frac{1}{3}$
  & $-\frac{1}{2}$ & $1$ & $0$ & $0$ & $0$ & $0$ & $\frac{1}{2}$
  & $-\frac{1}{2}$ & $\frac{1}{2}$ & $-\frac{1}{2}$ & $-\frac{1}{3}$
  & $\frac{1}{3}$ & $-\frac{1}{2}$ & $\frac{1}{2}$ \\
 \hline
 $\sqrt{{40}}Q^{N}_i$ & $1$ & $1$ & $2$ & $2$ & $1$ & $0$ & $5$ & $-5$ & $5$
  & $-2$ & $-3$ & $-2$ & $-3$ & $-2$ & $-3$ & $2$ & $-2$ \\
 \hline
 $\tilde{Z}_2^H$
  & - & - & - & - & - & - & + & + & - & + & + & - & - & - &  - &  + & + \\
 \hline
 $Z_2^M$
  & - & - & - & - & - & - & + & + & + & + & + & + & + & + &  + &  - & - \\
 \hline
 $Z_2^E$
  & + & + & + & + & + & + & + & + & - & + & + & - & - & - &  - &  - & - \\
 \hline
\end{tabular}
\caption{Representations of the chiral superfields under the $SU(3)$ and
$SU(2)$ gauge groups, and their $E_6$ normalized $U(1)_Y$ and $U(1)_N$ charges.
The transformation properties under the discrete symmetries $\tilde{Z}_2^H$,
$Z_2^M$ and $Z_2^E$ are also shown.  Note that we omit the pure gauge singlet,
$\hat{\phi}$, as it transforms trivially under all of the gauge and discrete
symmetries.}
\label{tab:charges}
\end{table*}

After imposing $\tilde{Z}_2^H$ symmetry, the low-energy superpotential
of the model can be written,
\begin{align}
W = {} & \, \lambda \hat{S} \hat{H}_d \cdot \hat{H}_u - \sigma
  \hat{\phi} \hat{S} \hat{\overline{S}} + \dfrac{\kappa}{3}\hat{\phi}^3
  + \dfrac{\mu}{2}\hat{\phi}^2 + \Lambda_F\hat{\phi} \nonumber \\
  & {} + \lambda_{\alpha\beta} \hat{S}
  \hat{H}^d_{\alpha} \cdot \hat{H}^u_{\beta} + \kappa_{ij} \hat{S}
  \hat{D}_{i} \hat{\overline{D}}_{j}
  + \tilde{f}_{i\alpha} \hat{S}_{i} \hat{H}_u \cdot \hat{H}^d_{\alpha}
  \nonumber \\
  & {} + f_{i\alpha} \hat{S}_{i} \hat{H}^u_{\alpha} \cdot \hat{H}_d
  + g^D_{ij} \hat{Q}_i \cdot \hat{L}_4 \hat{\overline{D}}_j \nonumber \\
  & {} + h^E_{i\alpha} \hat{e}^c_{i} \hat{H}^d_{\alpha} \cdot \hat{L}_4
  + \mu_L \hat{L}_4 \cdot \hat{\overline{L}}_4 \nonumber \\
  & {} + \tilde{\sigma} \hat{\phi} \hat{L}_4 \cdot \hat{\overline{L}}_4
  + W_{\text{MSSM}}(\mu=0)\,,
\label{eq:superpotential}
\end{align}
where $W_{\text{MSSM}}(\mu=0)$ is the MSSM superpotential without the $\mu$
term, all superfields appear with a hat and all
coefficients of the superfields are couplings of appropriate
dimensions, and $\hat{A}\cdot \hat{B} \equiv \epsilon_{\alpha \beta}
\hat{A}^\alpha \hat{B}^\beta = \hat{A}^2 \hat{B}^1 - \hat{A}^1 \hat{B}^2$.

The above superpotential interactions are supplemented by a set
of soft SUSY breaking interactions; namely, soft scalar masses
for all chiral superfields, soft breaking scalar trilinear, bilinear and linear
terms for each superpotential coupling, and soft breaking gaugino masses.  The
resulting large number of soft parameters can be substantially reduced by
considering constrained SUSY models inspired by gravity mediated
SUSY breaking.  Here we assume that at the GUT scale, $M_X$, all scalar masses
are unified to a common value $m_0$, all gaugino masses are unified to
$M_{1/2}$, that all soft trilinears are equal to $A_0$, and all soft bilinears
to $B_0$.

Once these soft mass parameters are evolved down to the EW
scale, minimizing the Higgs potential (given in
Ref.~\cite{Athron:2014pua}) leads to the Higgs fields developing VEVs,
\begin{equation}
\begin{aligned}
\langle H_d \rangle &= \frac{1}{\sqrt{2}} \begin{pmatrix} v_1 \\ 0
\end{pmatrix} , \quad \langle H_u \rangle = \frac{1}{\sqrt{2}} \begin{pmatrix}
0 \\ v_2 \end{pmatrix} , \\[2mm]
\langle S \rangle &= \frac{s_1}{\sqrt{2}} , \quad
\langle \overline{S} \rangle = \frac{s_2}{\sqrt{2}}, \quad
\langle \phi \rangle = \frac{\varphi}{\sqrt{2}} .
\end{aligned}
\label{eq:vevs}
\end{equation}

\section{Analysis and Results}
\label{sec:results}
To determine the sparticle spectrum we created a spectrum generator for the
model using \FSv \cite{Athron:2014yba,Athron:2014wta} coupled with \SARAHv
\cite{Staub:2008uz,Staub:2010jh,Staub:2012pb,Staub:2013tta}.  Internally \FS
also uses some routines from \SOFTSUSY \cite{Allanach:2001kg, Allanach:2013kza}.
We focused on the scenarios where the $Z^\prime$ mass is decoupled from the EWSB
conditions, and so choose the SM singlet VEVs to satisfy
$s = \sqrt{s_1^2 + s_2^2} = 650$ TeV, giving $m_{Z^\prime} \approx 240$ TeV.
At the same time, to reproduce the relic density with a Higgsino or
mixed bino-Higgsino DM candidate we looked for scenarios with a small
value of $\mu_{\text{eff}} = \lambda s_1 / \sqrt{2}$.  This implies that the
coupling $\lambda$ should be very small.  To find such solutions, we implemented
a new solver algorithm in \FS that makes use of semi-analytic solutions to the
renormalization group equations (RGEs).  This allows us to choose input values
of $\mu_{\text{eff}}$ and $M_{1/2}$ to obtain an acceptable DM
candidate.  For a given value of $A_0$ and $B_0$, the value of $m_0$ is then
fixed by the requirement of correct EWSB.  The remaining EWSB conditions are
used to fix the ratio $\tan\theta = s_2 / s_1$, the VEV $\varphi$, and the
superpotential linear coupling, $\Lambda_F$, as well as its soft breaking
counterpart $\Lambda_S$.  The full details of this procedure will be given in
our companion paper \cite{long-paper}.

For scenarios with such small values of $\lambda$ we find that setting the
exotic couplings $\kappa_i$ and $\lambda_\alpha$ to values much larger than
$\lambda$ induces large mixings amongst exotic scalars, leading to tachyons.
Therefore we choose these couplings to be of a similar size to $\lambda$.  For
simplicity we also set the remaining exotic couplings $g^D_{ij}$,
$h^E_{i\alpha}$, $\tilde{f}_{i\alpha}$ and $f_{i\alpha}$ to negligibly small
values.  The choice of small $\kappa_i$ and $\lambda_\alpha$ allows light
exotic fermions to be present in the spectrum.

However, the solutions we found also have many heavy states as well,
and in particular very heavy stops.  To obtain a precise prediction
for the lightest Higgs mass we used the effective field theory
approach of \SUSYHDv \cite{Vega:2015fna}.  To do this we performed a
tree-level matching to the MSSM at the scale $M_S =
\sqrt{m_{\tilde{t}_1}^{\text{\DRbar}} m_{\tilde{t}_2}^{\text{\DRbar}}}$,
e.g.~ by setting the MSSM soft scalar masses to be those obtained in the
CSE$_6$SSM after running from the GUT scale.  Since the exotic couplings
beyond the MSSM are very small, the associated logarithms in the Higgs mass
are negligible and so this approach should not degrade the accuracy of our
calculations.

The calculated particle spectrum of the model for six benchmark points is given
in \tabref{tab:spectrum}.  For all of the benchmarks the light Higgs mass is
consistent with the measured value \cite{Aad:2015zhl}, within theoretical
errors.

\begin{table*}[h!]
\centering
\small
  \begin{tabular}{ | c || c | c | c | c | c | c | }
    \hline
     & \textbf{BM 1} & \textbf{BM 2} & \textbf{BM 3} & \textbf{BM 4}
       & \textbf{BM 5} & \textbf{BM 6} \\
    \hline
    $\lambda(M_X)$ & $0.0009152$ & $0.0009886$ & $0.0007052$ & $0.002295$
      & $0.00047$ & $0.0005$ \\
    $\lambda_{1,2}(M_X)$ & $0.001$ & $0.0013$ & $0.0012$ & $0.003$ & $0.0016$
      & $0.0012$ \\
    $\kappa_{1,2,3}(M_X)$ & $0.001$ & $0.0013$ & $0.0012$ & $0.00135$
      & $0.0016$ & $0.0012$ \\
    $M_{1/2}$ $[\textrm{GeV}]$ & $2227.79$ & $2407.79$ & $1617.79$
      & $5800.98$ & $1900.00$ & $2017.79$ \\
    $m_0$ $[\textrm{GeV}]$ & $9586.46$ & $9494.22$ & $8800.16$
      & $1.084\cdot 10^{4}$ & $7396.89$ & $7410.12$ \\
    $A_0$ $[\textrm{GeV}]$ & $-7281.96$ & $-6481.96$ & $-7541.96$ & $2129.63$
      & $-4600.00$ & $-4441.96$ \\
    \hline
    $1 - \tan\theta$ & $1.5\cdot 10^{-6}$ & $1.9\cdot 10^{-6}$
      & $2.4\cdot 10^{-6}$ & $9.4\cdot 10^{-7}$ & $5.3\cdot 10^{-6}$
      & $2.7\cdot 10^{-6}$ \\
    $\varphi$ $[\textrm{TeV}]$ & $-1633$ & $-1493$ & $-1737$ & $-708$
      & $-1713$ & $-1621$ \\
    $\Lambda_F^{1/2}$ $[\textrm{TeV}]$ & $127$ & $120$ & $131$ & $108$
      & $139$ & $133$ \\
    $\Lambda_S^{1/3}$ $[\textrm{TeV}]$ & $98$ & $91$ & $102$ & $61$ & $101$
      & $96$ \\
    \hline
    $m_{\tilde{q}_{1,2}}$ $[\textrm{GeV}]$ & $9400$ & $9400$ & $8500$
      & $12500$ & $7300$ & $7350$ \\
    $m_{\tilde{l}}$ $[\textrm{GeV}]$ & $9500$ & $9400$ & $8700$ & $11000$
      & $7330$ & $7350$ \\
    $m_{\tilde{b}_1}$ $[\textrm{GeV}]$ & $7577$ & $7616$ & $6759$ & $10801$
      & $5927$ & $5992$ \\
    $m_{\tilde{b}_2}$ $[\textrm{GeV}]$ & $9361$ & $9364$ & $8438$ & $12411$
      & $7287$ & $7345$ \\
    $m_{\tilde{t}_1}$ $[\textrm{GeV}]$ & $5476$ & $5550$ & $4802$ & $8582$
      & $4326$ & $4396$ \\
    $m_{\tilde{t}_2}$ $[\textrm{GeV}]$ & $7580$ & $7619$ & $6762$ & $10803$
      & $5931$ & $5995$ \\
    \hline
    $m_{H^\pm} \approx m_{A_2} \approx m_{h_3}$ $[\textrm{GeV}]$ & $9381$
      & $9312$ & $8576$ & $11056$ & $7245$ & $7266$ \\
    $m_{A_1}$ $[\textrm{GeV}]$ & $5193$ & $6605$ & $2723$ & $9978$ & $931$
      & $3650$ \\
    $m_{A_3}$ $[\textrm{GeV}]$ & $42896$ & $39797$ & $44939$ & $25797$
      & $43985$ & $41946$ \\
    $m_{h_1}$ $[\textrm{GeV}]$ & $125.22$ & $125.04$ & $124.96$ & $125.04$
      & $124.04$ & $124.10$ \\
    $m_{h_2}$ $[\textrm{GeV}]$ & $8208$ & $8289$ & $7985$ & $8048$ & $7072$
      & $7195$ \\
    $m_{h_4}$ $[\textrm{GeV}]$ & $38770$ & $36136$ & $40469$ & $24529$
      & $39664$ & $37913$ \\
    $m_{Z^\prime} \approx m_{h_5}$ $[\textrm{GeV}]$ & $2.4\cdot 10^{5}$
      & $2.4\cdot 10^{5}$ & $2.4\cdot 10^{5}$ & $2.4\cdot 10^{5}$
      & $2.4\cdot 10^{5}$ & $2.4\cdot 10^{5}$ \\
    \hline
    $m_{\tilde{D}_1}(1,2,3)$ $[\textrm{GeV}]$ & $8523$ & $8430$ & $7016$
      & $12308$ & $4520$ & $5562$ \\
    $m_{\tilde{D}_2}(1,2,3)$ $[\textrm{GeV}]$ & $10376$ & $10516$ & $9966$
      & $12662$ & $9698$ & $9062$ \\
    $\mu_D(1,2,3)$ $[\textrm{GeV}]$ & $1243$ & $1575$ & $1499$ & $1540$
      & $1943$ & $1489$ \\
    \hline
    $m_{H_1^\pm}(1,2)$ $[\textrm{GeV}]$ & $8938$ & $8762$ & $7862$ & $10433$
      & $5799$ & $6309$ \\
    $m_{H_2^\pm}(1,2)$ $[\textrm{GeV}]$ & $10056$ & $10091$ & $9490$
      & $11986$ & $8696$ & $8328$ \\
    $m_{H_1}(1,2)$ $[\textrm{GeV}]$ & $13406$ & $13332$ & $12935$ & $14251$
      & $12123$ & $12189$ \\
    $m_{H_2}(1,2)$ $[\textrm{GeV}]$ & $17161$ & $17113$ & $16944$ & $17584$
      & $16560$ & $16494$ \\
    $\mu_{\tilde{H}^\pm}(1,2) \approx \mu_{\tilde{H}^0_{1,2}}(1,2)$
      $[\textrm{GeV}]$ & $580$ & $750$ & $700$ & $1663$ & $929$ & $699$ \\
    \hline
    $m_{S_{1,2,3}}$ $[\textrm{GeV}]$ & $25593$ & $25516$ & $25663$ & $24875$
      & $25583$ & $25567$ \\
    \hline
    $m_{L_{4,1}^\pm}$ $[\textrm{GeV}]$ & $17580$ & $17468$ & $17355$
      & $17512$ & $16663$ & $16657$ \\
    $m_{L_{4,2}^\pm}$ $[\textrm{GeV}]$ & $18465$ & $18422$ & $18021$
      & $19611$ & $17470$ & $17513$ \\
    $m_{L_{4,1}^0}$ $[\textrm{GeV}]$ & $19994$ & $19886$ & $19870$ & $19671$
      & $19345$ & $19336$ \\
    $m_{L_{4,2}^0}$ $[\textrm{GeV}]$ & $20771$ & $20724$ & $20449$ & $21557$
      & $20039$ & $20072$ \\
    $\mu_{\tilde{L}_4^\pm} \approx \mu_{\tilde{L}_{4,1}^0},\mu_{\tilde{L}_{4,2}^
      0}$ $[\textrm{GeV}]$ & $15358$ & $15314$ & $15439$ & $14955$ & $15436$
      & $15447$ \\
    \hline
  \end{tabular}
\caption{Parameters for the benchmark points BM1--BM6 and the resulting
sparticle masses.  For all points we fix $s = 650$ TeV,
$\tan\beta(M_Z) \equiv v_2 / v_1 = 10$, $\sigma(M_X) = 0.02$,
$\kappa(M_X) = 0.01$, $\mu(M_X) = 0$ GeV, $\mu_L(M_X) = 10$ TeV,
$\tilde{\sigma}(M_X) = 0$ and $B_0 = 0$ GeV.  The couplings
$\tilde{f}_{i\alpha}$, $f_{i\alpha}$, $h^E_{i\alpha}$ and $g^D_{ij}$
are all set to negligibly small values, as they do not have
a significant impact on the spectrum.  For brevity, we show an approximate mass
$m_{\tilde{q}_{1,2}}$ for the first and second generation up- and down-type
squarks.  The exact masses of all four states are within $\pm 100$ GeV of this
value.  Similarly, $m_{\tilde{l}}$ represents an approximate mass for all
sleptons, with the exact masses all lying within $\pm 150$ GeV of the given
value.}
\label{tab:spectrum}
\end{table*}

The altered RG flow in this model ensures that the sfermions are
heavier than the gauginos.  Additionally, the requirement of a light
$\mu_{\text{eff}}$ leads to the EWSB conditions imposing a
relationship amongst the universal soft masses such that $m_0 > A_0,
M_{1/2}$.  This means that maximal mixing in the stop sector cannot be
used to obtain a 125 GeV Higgs and so in all six benchmarks the
sfermions are very heavy and well beyond the reach of the LHC.

Conversely, the light exotic leptoquarks and inert Higgsinos that result from
the small exotic couplings can be detectable at the LHC.  The
leptoquark states, $D_i$, participate in QCD interactions and may be pair
produced at the LHC.  When past threshold the differential production cross
section is comparable to the pair production of top quarks \cite{Athron:2011wu}.
These states are $R$-parity odd and therefore decay with missing energy,
through a long cascade decay involving the couplings $g_{ij}^D$ in
Eq.~(\ref{eq:superpotential}) to allow the initial decay of the exotic
quark into a squark (quark) and exotic lepton (slepton) pair, and also
$h_{i\alpha}^E$ for the exotic lepton (or slepton), $L_4$, to decay.  Since
there is a hierarchy in the SM Yukawa couplings it seems natural to assume
that a similar hierarchy will exist in the leptoquark and $\hat{L}_4$ Yukawa
interactions.  In this case pair production will therefore give rise to an
enhancement of $pp \to t\bar{t}\tau^+\tau^- +E^{miss}_T + X$ and
$pp \to b\bar{b} + E_T^{miss} +X$, where $X$ stands for any number of light
quark/gluon jets.

The exotic charged and neutral inert Higgsino states may be produced
in pairs through off-shell $W$ and $Z$ bosons.  They subsequently decay
into an on-shell $Z$ or $W$ and a singlino from $f$- and $\tilde{f}$-coupling
induced mixing\footnote{They may also decay through the $f$- and
$\tilde{f}$-couplings into a Higgs boson and a singlino state.}.  Thus the
presence of these states at very low energies should enhance
$pp \to ZZ + E_T^{miss} + X$, $pp \to WZ + E_T^{miss} + X$ and
$pp \to WW + E_T^{miss} + X$.  Note that this signature differs from the one
which has been considered in previous $E_6$ constructions, where they decayed
into fermion-sfermion pairs via couplings that are forbidden in this model
by the $\tilde{Z}_2^H$ symmetry.

Although the sfermions are rather heavy, in all benchmark points other
than BM4 the MSSM-like neutralinos and charginos are also light in
addition to the exotic states.  The neutralino and chargino masses
are shown in \tabref{tab:dark-matter}.  While these are weakly
interacting states, they are very light, so it is reasonable to expect
some discovery potential, in particular from the production of a
neutralino-chargino pair, which leads to an enhancement of $pp \to lll
+ E_T^{miss} + X$.  The branching ratios for the processes $\tilde{\chi}_2^0
\to \tilde{\chi}_1^0 l \bar{l}$ and $\tilde{\chi}_1^\pm \to \tilde{\chi}_1^0
l \nu_l$, obtained using a \CalcHEP \cite{Belyaev:2012qa} model generated
using \SARAHv, are shown in \tabref{tab:dark-matter}.  For the scenarios
considered here, the process $\tilde{\chi}_2^0 \to \tilde{\chi}_1^0 l \bar{l}$
proceeds almost entirely through diagrams involving a virtual $Z$, with
diagrams involving a virtual Higgs being a negligible contribution due to the
small mass splitting between $m_{\tilde{\chi}_2^0}$ and $m_{\tilde{\chi}_1^0}$
and the small Higgs couplings to leptons and quarks\footnote{Note that the
decay of  $\tilde{\chi}_2^0$ into $ \tilde{\chi}_1^0 + t \bar{t}$ is not
kinematically allowed.}.  Therefore the discovery prospects
are expected to be rather similar to those in the $WZ$-mediated scenario of
Ref.~\cite{runii-neutralinos-3lepton}.

Currently, stronger constraints
can be placed on the gaugino sector by the measurement of the relic
density of DM and limits on the spin independent (SI) cross section
from direct detection experiments.  The composition of
the lightest neutralino, relic density along with a breakdown of
the various contributions to the annihilation cross section and the SI
and spin dependent cross sections are also given in \tabref{tab:dark-matter}.
To calculate DM observables in the model, the generated \CalcHEP
model files were used to implement the model in \MICROMEGASv
\cite{Belanger:2001fz,Belanger:2004yn, Belanger:2006is,Belanger:2008sj,
  Belanger:2010gh,Belanger:2013oya, Belanger:2014vza}.  The inert singlinos
are almost massless and so have a negligible contribution to the total relic
density.  The total relic density shown is that due to the lightest neutralino.

To obtain the observed relic density \cite{Ade:2015xua} one may use a pure
Higgsino DM candidate with a mass of about $1$ TeV.  However this then
requires a very heavy bino to ensure the lightest neutralino is pure
Higgsino and that in turn means the gluino must be above the reach of
LHC run II in this constrained model.  BM4 is an example of such a scenario.
In this scenario the SI cross section is reasonably far from the
current best exclusion limit of LUX \cite{Akerib:2013tjd}, though \xenon will
be in a position to either discover this or rule it out.

The SI cross section increases in scenarios where the lightest
neutralino is a mixture of bino and Higgsino.  In such cases the SI
cross section is very close to the LUX limit\footnote{In fact while
  this document was in preparation a reinterpretation of the LUX
  limits appeared on the arXiv pre-print server \cite{Akerib:2015rjg},
  which makes the tension more severe.  However despite this tension
  it is still possible that points like these could be discovered by
  \xenon and therefore they remain very interesting.} and will be
discoverable in the ``early data'' of \xenon.  BMs 1-3 are examples of
this.  In this case the correct relic density is achieved with a much
lighter DM candidate and subsequently the gluino is within reach of
the LHC and gluino pair production will lead to a considerable
enhancement of $pp \to q\overline{q}q\overline{q} +E_T^{miss} +
X$.  BM1 has exotic leptoquarks with masses below current limits on the
gluino and should be easily discoverable at the LHC run II, while for
BM2 the exotic quarks are now heavier but should still be within the
reach of the LHC run II.  In both BM1 and BM2 the gluino mass is
fairly large though the LHC should still be able to discover
them, at least with the high luminosity upgrade \cite{Baer:2012vr}.
In BM3 both the gluino and the leptoquarks are very light and
discovery of these should be possible with $300$ fb$^{-1}$ of
integrated luminosity (IL).

Finally, BM5 and BM6 represent scenarios with a Higgsino DM candidate
that is too light to account for all of the observed DM relic density.
This substantially decreases the direct detection event rate, allowing
the LUX cross section limits to be evaded and reducing the sensitivity
of \xenon to these points.  At the same time, both the gluino and
exotic quark masses are light enough to be accessible at run II.  In
contrast to BMs 1-3, these points could therefore be discovered at run
II of the LHC, without being in tension with the current LUX limits or
being observed in the early \xenon data.  However, this comes at the
cost of requiring an additional source of DM in this scenario in order
to explain the observed relic density.  BM5 also shows that the
leptoquarks can be heavier than in the other benchmarks so that it may
be challenging to find with $300$ fb$^{-1}$ of IL, but it
still should be possible to discover these at the LHC.  In contrast in BM6 the
leptoquark is comparatively light but the gluino may require longer running to
be discovered.

\begin{table*}[ht]
\centering
\small
  \begin{tabular}{ | c || c | c | c | c | c | c | }
    \hline
     & \textbf{BM 1} & \textbf{BM 2} & \textbf{BM 3} & \textbf{BM 4}
       & \textbf{BM 5} & \textbf{BM 6} \\
    \hline
    $m_{\tilde{g}}$ $[\textrm{GeV}]$ & $2099$ & $2256$ & $1541$ & $5230$
      & $1716$ & $1839$ \\
    \hline
    $m_{\tilde{\chi}_1^\pm}$ $[\textrm{GeV}]$ & $422$ & $454$ & $320$
      & $1034$ & $216$ & $231$ \\
    $m_{\tilde{\chi}_2^\pm} \approx m_{\tilde{\chi}_4^0}$ $[\textrm{GeV}]$
      & $780$ & $845$ & $570$ & $2129$ & $645$ & $682$ \\
    $m_{\tilde{\chi}_1^0}$ $[\textrm{GeV}]$ & $375$ & $409$ & $264$ & $1024$
      & $204$ & $219$ \\
    $m_{\tilde{\chi}_2^0}$ $[\textrm{GeV}]$ & $433$ & $464$ & $338$ & $1038$
      & $226$ & $241$ \\
    $m_{\tilde{\chi}_3^0}$ $[\textrm{GeV}]$ & $445$ & $479$ & $338$ & $1159$
      & $336$ & $358$ \\
    $m_{\tilde{\chi}_5^0}$ $[\textrm{GeV}]$ & $25394$ & $23602$ & $26745$
      & $14546$ & $26437$ & $25249$ \\
    $m_{\tilde{\chi}_6^0}$ $[\textrm{GeV}]$ & $29853$ & $27651$ & $31546$
      & $16364$ & $31173$ & $29737$ \\
    $m_{\tilde{\chi}_7^0}$ $[\textrm{GeV}]$ & $231028$ & $232102$ & $230097$
      & $238639$ & $230406$ & $231254$ \\
    $m_{\tilde{\chi}_8^0}$ $[\textrm{GeV}]$ & $258656$ & $257259$ & $259681$
      & $249541$ & $259532$ & $258784$ \\
    \hline
    $|(Z_N)_{11}|^2$ & $0.6318$ & $0.6075$ & $0.7210$ & $0.0691$ & $0.0679$
      & $0.0624$ \\
    $|(Z_N)_{12}|^2$ & $0.0081$ & $0.0075$ & $0.0106$ & $0.0028$ & $0.0180$
      & $0.0165$ \\
    $|(Z_N)_{13}|^2 + |(Z_N)_{14}|^2$ & $0.3601$ & $0.3850$ & $0.2685$
      & $0.9281$ & $0.9141$ & $0.9211$ \\
    \hline
    $\textrm{BR}(\tilde{\chi}_1^- \to \tilde{\chi}_1^0 l \bar{\nu}_l)$
      & $0.2220$ & $0.2220$ & $0.2220$ & $0.2280$ & $0.2260$ & $0.2260$ \\
    $\text{BR}( \tilde{\chi}_2^0 \to \tilde{\chi}_1^0 l \bar{l}\,)$
      & $0.0689$ & $0.0689$ & $0.0684$ & $0.0733$ & $0.0670$ & $0.0674$ \\
    \hline
    $\Omega h^2$ & $0.1188$ & $0.1185$ & $0.1187$ & $0.1184$ & $0.01055$
      & $0.009626$ \\
    \hline
    $\sigma_{\mathrm{SI}}^{\mathrm{p}}$ $[\times 10^{-45}\textrm{ cm}^2]$
      & $5.88$ & $6.14$ & $4.84$ & $2.35$ & $4.67$ & $4.32$ \\
    $\sigma_{\mathrm{SD}}^{\mathrm{p}}$ $[\times 10^{-41}\textrm{ cm}^2]$
      & $6.4$ & $5.58$ & $10.0$ & $0.3529$ & $15.8$ & $12.8$ \\
    $\sigma_{\mathrm{SI}}^{\mathrm{n}}$ $[\times 10^{-45}\textrm{ cm}^2]$
      & $5.97$ & $6.24$ & $4.91$ & $2.39$ & $4.75$ & $4.39$ \\
    $\sigma_{\mathrm{SD}}^{\mathrm{n}}$ $[\times 10^{-41}\textrm{ cm}^2]$
      & $4.9$ & $4.27$ & $7.66$ & $0.2699$ & $12.1$ & $9.78$ \\
    \hline
    $\tilde{\chi}_1^0 \tilde{\chi}_1^0 \to t \bar{t}$ (\%) & $44.9$ & $39.0$
      & $60.0$ & $0.6$ & $0.5$ & $3.3$  \\
    $\tilde{\chi}_1^0 \tilde{\chi}_1^0 \to W^+ W^-$ (\%) & $20.6$ & $19.4$
      & $21.6$ & $5.0$ & $27.9$ & $22.0$ \\
    $\tilde{\chi}_1^0 \tilde{\chi}_1^0 \to Z Z$ (\%) & $13.2$ & $12.8$
      & $11.4$ & $3.9$ & $18.4$ & $14.1$ \\
    $\tilde{\chi}_1^0 \tilde{\chi}_1^0 \to Z h_1$ (\%) & $2.9$ & $2.7$
      & $2.9$ & $0.7$ & $0.0$ & $1.7$ \\
    $\tilde{\chi}_1^0 \tilde{\chi}_1^0 \to h_1 h_1$ (\%) & $0.5$ & $0.4$
      & $0.9$ & $0.02$ & $0.1$ & $0.1$ \\
    $\tilde{\chi}_1^0 \tilde{\chi}_1^- \to W^- Z$ (\%) & $0.8$ & $1.1$
      & $0.2$ & $1.4$ & $0.2$ & $1.4$ \\
    $\tilde{\chi}_1^0 \tilde{\chi}_1^- \to W^- h_1$ (\%) & $1.3$ & $1.6$
      & $0.3$ & $1.5$ & $2.7$ & $2.6$ \\
    $\tilde{\chi}_1^+ \tilde{\chi}_1^- \to W^+ W^-$ (\%) & $0.1$ & $0.1$
      & $2 \cdot 10^{-3}$ & $1.9$ & $0.5$ & $0.7$ \\
    $\tilde{\chi}_1^0 \tilde{\chi}_1^- \to \gamma W^-$ (\%) & $0.6$ & $0.8$
      & $0.1$ & $1.4$ & $1.5$ & $1.5$ \\
    $\tilde{\chi}_1^0 \tilde{\chi}_1^- \to d_i \bar{u}_i$ (\%) & $8.8$ & $12.0$
      & $1.6$ & $25.7$ & $29.4$ & $30.0$ \\
    $\tilde{\chi}_1^0 \tilde{\chi}_1^- \to l_i^- \bar{\nu}_{l_i}$ (\%) & $2.7$
      & $3.8$ & $0.5$ & $8.8$ & $10.7$ & $10.8$ \\
    $\tilde{\chi}_2^0 \tilde{\chi}_1^- \to d_i \bar{u}_i$ (\%) & $0.2$ & $0.4$
      & $3 \cdot 10^{-3}$ & $12.0$ & $0.7$ & $1.2$ \\
    $\tilde{\chi}_1^0 \tilde{\chi}_2^0 \to d_i \bar{d}_i$ (\%) & $0.9$ & $1.4$
      & $0.07$ & $6.4$ & $1.5$ & $2.0$ \\
    $\tilde{\chi}_1^0 \tilde{\chi}_2^0 \to u_i \bar{u}_i$ (\%) & $0.8$ & $1.3$
      & $0.06$ & $4.7$ & $0.9$ & $1.3$ \\
    $\tilde{\chi}_1^+ \tilde{\chi}_1^- \to d_i \bar{d}_i$ (\%) & $0.1$ & $0.2$
      & $4 \cdot 10^{-3}$ & $3.0$ & $0.9$ & $1.2$ \\
    $\tilde{\chi}_1^+ \tilde{\chi}_1^- \to u_i \bar{u}_i$ (\%) & $0.2$ & $0.3$
      & $6 \cdot 10^{-3}$ & $4.9$ & $1.1$ & $1.6$ \\
    $\tilde{\chi}_2^0 \tilde{\chi}_1^- \to l^-_i \bar{\nu}_{l_i}$ (\%) & $0.1$
      & $0.1$ & $9 \cdot 10^{-4}$ & $4.1$ & $0.2$ & $0.4$ \\
    \hline
  \end{tabular}
\caption{Masses of the charginos and neutralinos, the bino, wino and
higgsino components of the lightest neutralino ($|(Z_N)_{11}|^2$,
$|(Z_N)_{12}|^2$ and $|(Z_N)_{13}|^2 + |(Z_N)_{14}|^2$, respectively),
the branching ratios for the decays $\tilde{\chi}_1^- \to \tilde{\chi}_1^0
l \bar{\nu}_l$, $\tilde{\chi}_2^0 \to \tilde{\chi}_1^0 l \bar{l}$ (with
$l = e, \mu$) and the predicted relic density and WIMP-nucleon
scattering cross sections for the benchmark points BM1--6.  Also shown are
the approximate percentage contributions to the annihilation cross section
from the indicated channels for each benchmark.  Note that the contributions
to the total relic density are computed using the freeze-out approximation.}
\label{tab:dark-matter}
\end{table*}

\section{Conclusions}
\label{sec:conclusions}
In this letter we have presented benchmark scenarios in a new well
motivated $E_6$ inspired model, all of which predict states which can
be discovered at both \xenon and run II of the LHC.  With initial run II
results already available and new results from \xenon expected very soon
these scenarios are of urgent interest.

In BMs 1-3 we show that the model can explain DM, fitting the
observed relic density, while having exotic leptoquarks, gluinos, and
possibly even neutralinos and charginos discoverable at the LHC run II.
Further the bino-Higgsino DM candidate for these these points
should be discovered immediately in ``early data'' from the \xenon
experiment.

BM4 on the other hand shows a Higgsino dominated DM candidate, where
mixing with the bino is suppressed as the bino is rather heavy.  In
this case gaugino mass universality and the RG flow make the gluino
far too heavy for the LHC reach.  However the model can still be
discovered through exotic leptoquarks.  This emphasizes the need for
dedicated studies on these exotic states.  The DM is still within
discovery range of \xenon but should take a little longer to discover
than the other benchmarks.

Finally we also presented BM5 and BM6 where we showed that one can also have
scenarios with light phenomenology within reach of the LHC, where the relic
density is not fully explained.  In such a case the sensitivity of
\xenon will be limited by the substantially reduced relic density
for the lightest neutralino.  However even in this case the state
ought to be discoverable by the end of the \xenon experiment.

\section*{Acknowledgements}
This work was supported by the University of Adelaide and the Australian
Research Council through the ARC Centre of Excellence for Particle Physics
at the Terascale.

\section*{\refname}
\bibliography{bibliography}

\end{document}